\documentclass[twocolumn,aps,prd,superscriptaddress,nofootinbib,showpacs]{revtex4-2}

\usepackage{amsmath}
\usepackage{graphicx}
\usepackage{subcaption}
\usepackage{slashed}
\usepackage[colorlinks=true, allcolors=blue]{hyperref}

\begin{document}

\title{Two-loop QCD-corrections to $e^{+} e^{-} \rightarrow Z^{\ast} \rightarrow \boldsymbol{J /\psi}+\boldsymbol{J/ \psi}$ }

\author{Xiang Chen}
\email{xiang.chen@physik.uzh.ch}
\affiliation{Physik-Institut, Universit\"{a}t Z\"{u}rich, Winterthurerstrasse 190, CH-8057 Z\"{u}rich, Switzerland}

\author{Xin Guan}
\email{guanxin@slac.stanford.edu}
\affiliation{SLAC National Accelerator Laboratory, Stanford University, Stanford, CA 94039, USA}

\author{Chuan-Qi He}
\email{legend\_he@m.scnu.edu.cn}
\affiliation{State Key Laboratory of Nuclear Physics and Technology, Institute of Quantum Matter, South China Normal University, Guangzhou 510006, China}
\affiliation{Guangdong Basic Research Center of Excellence for Structure and Fundamental Interactions of Matter, Guangdong Provincial Key Laboratory of Nuclear Science, Guangzhou 510006, China}
\affiliation{Department of Physics and Astronomy, University of California, Los Angeles, CA 90095, USA}

\author{Yan-Qing Ma}
\email{yqma@pku.edu.cn}
\affiliation{School of Physics, Peking University, Beijing 100871, China}
\affiliation{Center for High Energy Physics, Peking University, Beijing 100871, China}

\begin{abstract}
We present a next-to-next-to-leading-order (NNLO) calculation within the nonrelativistic QCD (NRQCD) framework for the process \(e^{+}e^{-} \rightarrow Z^{\ast} \rightarrow J/\psi + J/\psi\). We find that the NNLO contribution is 2–3 times larger than the next-to-leading-order (NLO) contribution, which itself is already 3–5 times larger than the leading-order (LO) result. In the high-energy limit, we provide analytic expressions for the leading-power coefficients in the asymptotic expansion of the two-loop amplitudes. Our results are directly applicable to the bottomonium process \(Z \rightarrow \Upsilon + \Upsilon\). Using the obtained hadronic amplitudes, we predict the decay width of the \(Z\) boson into these rare di-charmonium and di-bottomonium final states.

\end{abstract}

\maketitle

\section{Introduction}

The study of exclusive double charmonium production in electron-positron annihilation has been a topic of intense theoretical and experimental interest since its first observation at the B factories in the early 2000s \cite{Belle:2002tfa,Belle:2004abn,Belle:2005lik,BaBar:2005nic}. Among these processes, $e^{+}e^{-} \to J/\psi + \eta_c$ became particularly notable for exhibiting one of the most significant discrepancies between experimental measurements and Standard Model predictions within the Non-Relativistic QCD (NRQCD) framework \cite{QuarkoniumWorkingGroup:2004kpm}, spurring extensive theoretical investigations \cite{Zhang:2005cha,He:2007te,Wang:2011qg,Jiang:2018wmv,Feng:2019zmt,Huang:2022dfw,Sun:2021tma,Sang:2022kub}. Such hard exclusive reactions provide a unique testing ground for probing the interplay between perturbative and non-perturbative QCD dynamics in heavy quarkonium physics.

A closely related process of considerable interest is the production of $J/\psi$-pairs. At B-factory energies, this configuration cannot proceed through a single virtual photon due to C-parity conservation. Instead, the dominant mechanism is the double-photon channel, whose leading-order (LO) cross section is sizable  \cite{Bodwin:2002fk,Bodwin:2002kk,Sang:2023nvt,Huang:2023pmn}. However, next-to-leading order (NLO) QCD corrections were found to be large and negative, with a K-factor ($\sigma_{\text{NLO}}/\sigma_{\text{LO}}$) ranging from -0.31 to 0.25 \cite{Gong:2007db}, providing a natural explanation for the non-observation of double $J/\psi$ production at these facilities. More recently, theoretical calculations have been extended to next-to-next-to-leading order (NNLO) within the framework of the vector dominance model \cite{Sang:2023nvt,Huang:2023pmn}, highlighting the dominance of the fragmentation-type contributions.
\begin{figure}[htb]
	\centering
    \begin{minipage}[b]{.45\linewidth}
        \centering
        \includegraphics[width=1.0\linewidth]{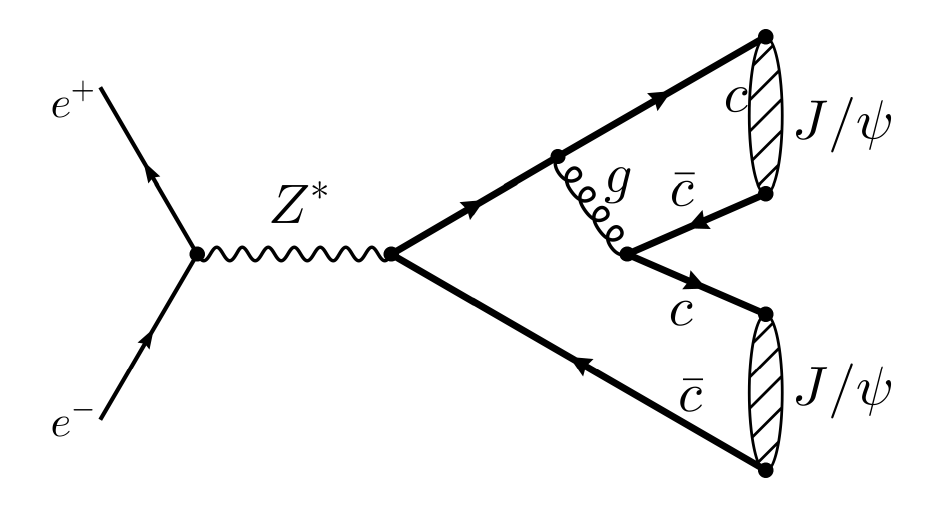}
    \subcaption{}
    \end{minipage}
    \begin{minipage}[b]{.45\linewidth}
        \centering
        \includegraphics[width=1.0\linewidth]{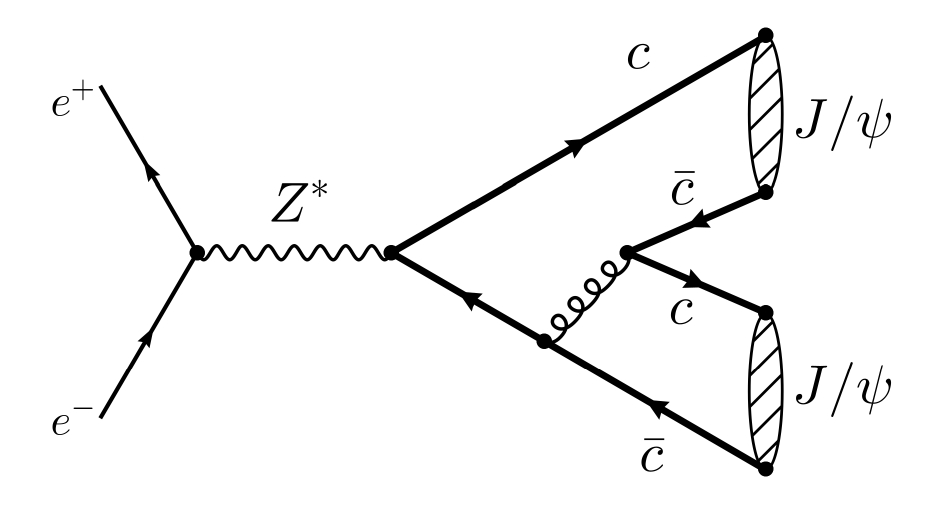}
    \subcaption{}
    \end{minipage}
    \begin{minipage}[b]{.45\linewidth}
        \centering
        \includegraphics[width=1.0\linewidth]{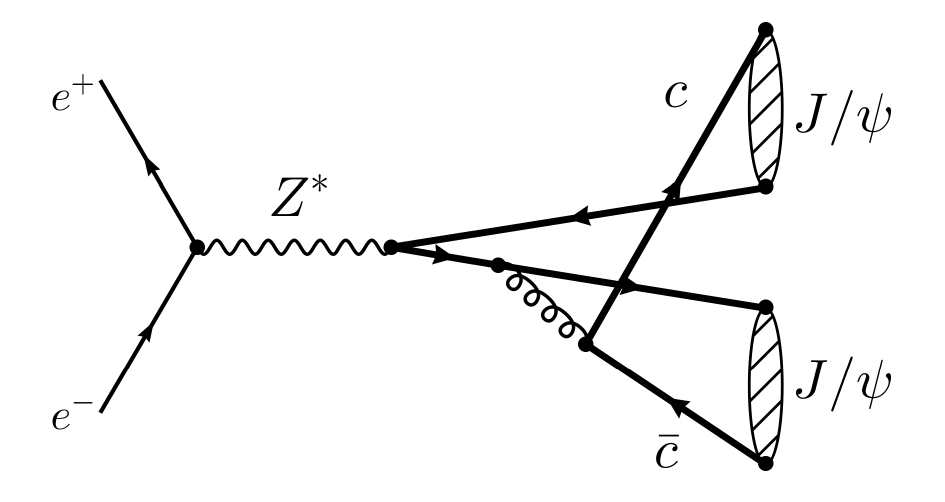}
    \subcaption{}
    \end{minipage}
    \begin{minipage}[b]{.45\linewidth}
        \centering
        \includegraphics[width=1.0\linewidth]{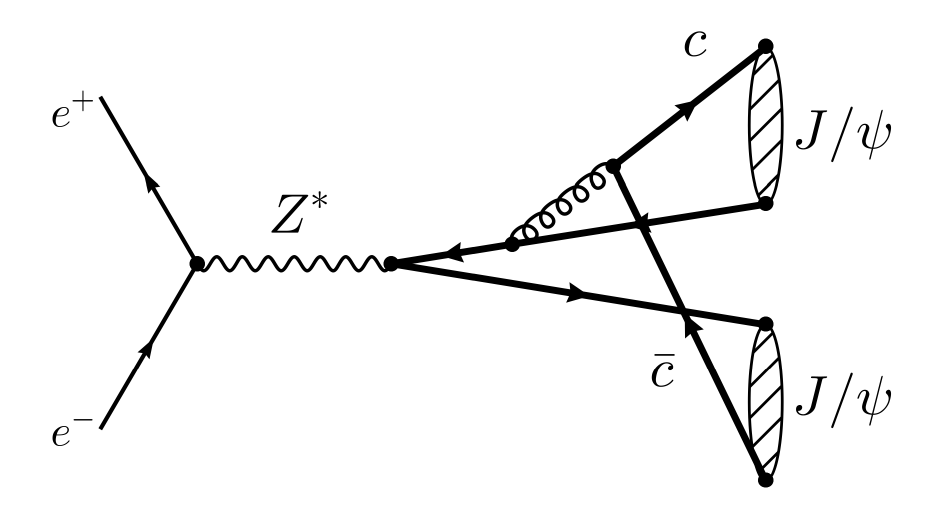}
    \subcaption{}
    \end{minipage}
    \caption{Tree diagrams for $e^{+} e^{-} \rightarrow Z^{\ast} \rightarrow J/\psi+J/\psi$.}\label{fig:ZJJtree}
\end{figure}

The experimental landscape has since evolved dramatically. The Belle II collaboration is ushering in a new era of precision in heavy flavor physics \cite{Belle-II:2018jsg}, while the LHC experiments have opened a new frontier by probing quarkonium production in Z-boson decays. Following a CMS proposal to search for Z decays into double quarkonia \cite{CMS:2019wch}, and motivated by the LHCb observation of novel structures in the $J/\psi$-pair mass spectrum \cite{LHCb:2020bwg}, the CMS collaboration has performed dedicated searches for the rare decay $Z \to J/\psi + J/\psi$. Using the full Run 2 dataset, they established a stringent upper limit on the branching fraction \cite{CMS:2022fsq}:
\begin{equation}
	\mathcal{B}(Z \to J/\psi + J/\psi) < 1.4 \times 10^{-6}.
\end{equation}

This new level of experimental precision demands equally precise theoretical predictions. It is well-established that NNLO QCD corrections within the NRQCD framework are often crucial for achieving accurate descriptions of quarkonium production processes \cite{Abreu:2022cco,Sang:2023hjl}. In this paper, we present a state-of-the-art calculation of the $Z \to J/\psi + J/\psi$ decay rate at NNLO in $\alpha_s$. Specifically, we provide the first complete computation of the two-loop amplitudes for this process. The paper is organized as follows. We begin by outlining the theoretical framework and establishing our notation. We then detail the calculation of the two-loop amplitudes, including the treatment of ultraviolet and infrared divergences through renormalization and factorization. The PSLQ algorithm~\cite{ferguson1998polynomial} is employed to reconstruct analytical expressions from high-precision numerical results. Finally, we provide a semi-analytical analysis of the decay width and conclude with a discussion of our findings and their implications.

\section{Framework and Methods}
We consider the two-loop QCD corrections to the following process
\begin{equation}\label{equ:process}
    e^+(k_1)+e^-(k_2)\rightarrow  Z^{\ast} \rightarrow J / \psi(P_1)+J / \psi(P_2),
\end{equation}
where the external momentum satisfy the on-shell conditions, 
\begin{equation}\label{equ:OnShell}
\begin{aligned}
& k_1^2=k_2^2=0, \\
& P_1^2=P_2^2=4m_c^2, \\
& s=q^2=(k_1+k_2)^2=(P_1+P_2)^2.
\end{aligned}
\end{equation}
Fig. \ref{fig:ZJJtree} presents the four tree diagrams.

Within the NRQCD formalism, the differential cross section for the process can be written as
\begin{equation}
d \sigma=d \hat{\sigma}_{e^{+} e^{-} \rightarrow c \bar{c}[n_1]+c \bar{c}[n_1]}\langle\mathcal{O}^{J / \psi}(n_1)\rangle^2,
\end{equation}
where the $d \hat{\sigma}$ denotes the short-distance partonic cross section for producing two $c\bar{c}$ pairs in intermediate states labeled by $n_1$, and the NRQCD long-distance matrix element(LDME) $\langle\mathcal{O}^{J / \psi}(n_1)\rangle$ describes the probability for these pairs to hadronize into physical $J / \psi$ mesons. For the $J /\psi$ production, we project the $c \bar{c}$ pair onto the color-singlet $n_1 = { }^3 S_1^{[1]}$, which corresponds to the quantum numbers of the physical meson. 

We define the amplitude of the process $Z^{\ast} \rightarrow J / \psi+ J / \psi$ as 
\begin{equation}\label{eq:ME}
\begin{aligned}
&\mathcal{A}^{\mu}(s)=\left\langle J / \psi(P_1)+ J / \psi(P_2)\left|J_{Z}^{\mu}\right| 0\right\rangle \\
=& \mathcal{A}^{\alpha_{1} \alpha_{2} \mu} (P_1, P_2)\, \varepsilon_{\alpha_{1}} \varepsilon_{\alpha_{2}}\\
=&\sum_{i=0} \sum^{6}_{j=1} g_{j}^{(i l)} \frac{e}{c_W s_W} \frac{1}{192\pi m_c^4} g_s^{2i+2} \\
&\times b_j^{\alpha_{1} \alpha_{2} \mu} \varepsilon_{\alpha_{1}} \varepsilon_{\alpha_{2}} M_{J/\psi} R^2_{J/\psi}(0). 
\end{aligned}
\end{equation}
It is expressed as a sum over Lorentz structures $b_j^{\alpha_1 \alpha_2 \mu}$, each multiplied by a scalar form factor $g_j^{(il)}(r)$, where $i$ denotes the loop order. The $\varepsilon_{\alpha_{1}}$ and $\varepsilon_{\alpha_{2}}$ denote the polarization vectors of the mesons. The factors $M_{J/\psi}$ and $R_{J/\psi}(0)$ represent the meson mass and the radial wave function at the origin, respectively. The meson mass can be approximated by $M_{J/\psi} = 2 m_c$.  The $R_{J/\psi}(0)$ encodes the nonperturbative transition amplitude of a color-singlet ${ }^3 S_1^{[1]}$ $c \bar{c}$ pair into a physical $J/\psi$, and its square is the corresponding NRQCD long-distance matrix element, as: $\langle\mathcal{O}^{J / \psi}({ }^3 S_1^{[1]})\rangle = |R_{J/\psi}(0)|^2$.

The interaction of the $Z$ boson with quarks involves an axial–vector current, with the presence of a $\gamma^5$ matrix. In dimensional regularization, $\gamma^5$ cannot be consistently defined with all its four-dimensional properties. In this work, we adopt the Larin scheme \cite{Larin:1993tq} to define $\gamma^5$,
\begin{equation}
\gamma^\mu \gamma^5 \rightarrow \frac{1}{3 !} \varepsilon^{\mu \nu \rho \sigma} \gamma_{\nu} \gamma_\rho \gamma_{\sigma}.
\end{equation}
As a cross-check, we also use the cyclic scheme, without commuting  $\gamma^5$ with any other gamma matrices. Both schemes lead to identical results after completing the Lorentz tensor reduction.

There are 8 possible Lorentz structures in total: 
\begin{equation}\label{eq:Lor}
\begin{aligned}
&b_1^{\alpha_{1} \alpha_{2} \mu}=\epsilon^{\alpha_1 \alpha_2 \mu P_1}, \\
&b_2^{\alpha_{1} \alpha_{2} \mu}=\epsilon^{\alpha_1 \alpha_2 \mu P_2}, \\
&b_3^{\alpha_{1} \alpha_{2} \mu}=\epsilon^{\alpha_2 \mu P_1 P_2}  P_{1}^{\alpha_1}/m_c^2, \\
&b_4^{\alpha_{1} \alpha_{2} \mu}=\epsilon^{\alpha_2 \mu P_1 P_2}  P_{2}^{\alpha_1}/m_c^2, \\
&b_5^{\alpha_{1} \alpha_{2} \mu}=\epsilon^{\alpha_1 \mu P_1 P_2}  P_{1}^{\alpha_2}/m_c^2, \\
&b_6^{\alpha_{1} \alpha_{2} \mu}=\epsilon^{\alpha_1 \mu P_1 P_2}  P_{2}^{\alpha_2}/m_c^2, \\
&b_7^{\alpha_{1} \alpha_{2} \mu}=\epsilon^{\alpha_1 \alpha_2 P_1 P_2}  P_{1}^{\mu}/m_c^2, \\
&b_8^{\alpha_{1} \alpha_{2} \mu}=\epsilon^{\alpha_1 \alpha_2 P_1 P_2}  P_{2}^{\mu}/m_c^2.
\end{aligned}
\end{equation}
Due to the identical nature of the two $J/\psi$, the amplitude has the relation $\mathcal{A}^{\alpha_{1} \alpha_{2} \mu} (P_1, P_2)=\mathcal{A}^{\alpha_{2} \alpha_{1} \mu} (P_2, P_1)$, leading to
\begin{equation}\label{eq:P1P2}
\begin{aligned}
&g_{1}^{(i l)} \equiv -g_{2}^{(i l)},\\
&g_{3}^{(i l)} \equiv -g_{6}^{(i l)},\\
&g_{4}^{(i l)} \equiv -g_{5}^{(i l)},\\
&g_{7}^{(i l)} \equiv  g_{8}^{(i l)}.
\end{aligned}
\end{equation}
The definition of a ${ }^3 S_1^{[1]}$ $c \bar{c}$ state requires $P_{1\,\alpha_1} \mathcal{A}^{\alpha_{1} \alpha_{2} \mu} =0$, resulting in
\begin{equation}\label{eq:GI}
\begin{aligned}
&g_{1}^{(i l)} =\frac{P_1^2}{m_c^2} g_{3}^{(i l)}+\frac{P_1 \cdot P_2}{m_c^2} g_{4}^{(i l)}.
\end{aligned}
\end{equation}
Furthermore, within the $\gamma^5$ scheme adopted in this work, the coefficient $g_{7}^{(i l)}$ vanishes:
\begin{equation}\label{eq:GI2}
g_{7}^{(i l)} = 0.
\end{equation}
The reason for this is tied to the Dirac structure of the amplitude. After loop integration and Lorentz contraction, the only gamma matrix structures that can appear are $\gamma^\mu \gamma^5$, $\gamma^{\alpha_1}$, $\gamma^{\alpha_2}$, $\slashed{P}_1$, and $\slashed{P}_2$. The tensor structure associated with $g_{7}^{(i l)}$ (and $g_{8}^{(i l)}$) in Eq.~\eqref{eq:Lor} cannot be constructed from this limited set, forcing its coefficient to be zero.

The Eqs.\eqref{eq:P1P2}, \eqref{eq:GI}  and \eqref{eq:GI2} reduce the amplitudes to only two independent coefficients at each order in $\alpha_s$; we choose $g_{1}^{(il)}$ and $g_{3}^{(il)}$ as the basis.

With the amplitude at hand, the total cross section for $e^{+} e^{-} \rightarrow Z^{\ast} \rightarrow J / \psi+J / \psi$ can be written as
\begin{equation}\label{eq:CS}
\begin{aligned}
\sigma=&\frac{\alpha(8 s_{W}^4-4 s_{W}^2+1)  }{384 c_{W}^2 s_{W}^2(m_{Z}^4+m_{Z}^2(\Gamma^2-2 s)+s^2)} \sqrt{1-\frac{16 m_c^2}{s}}  \\
&\times (\frac{q^{\mu}q^{\mu^{\prime}}} {q^2}-g^{\mu \mu^{\prime}} ) \mathcal{A}_{\mu}\mathcal{A}_{\mu^{\prime}}^{*}.
\end{aligned}
\end{equation}
We also study the partial decay width for Z boson in this channel, which is
\begin{equation}\label{eq:DW}
\Gamma= \frac{1}{2} \frac{1}{3} \frac{1}{16 \pi m_Z} \sqrt{1-\frac{16 m_c^2}{s}} (\frac{q^{\mu}q^{\mu^{\prime}}} {m_Z^2}-g^{\mu \mu^{\prime}} ) \mathcal{A}_{\mu}\mathcal{A}_{\mu^{\prime}}^{*}.
\end{equation}

\begin{figure*}[htb]
	\centering
    \begin{minipage}[b]{.3\linewidth}
        \centering
        \includegraphics[width=1.0\linewidth]{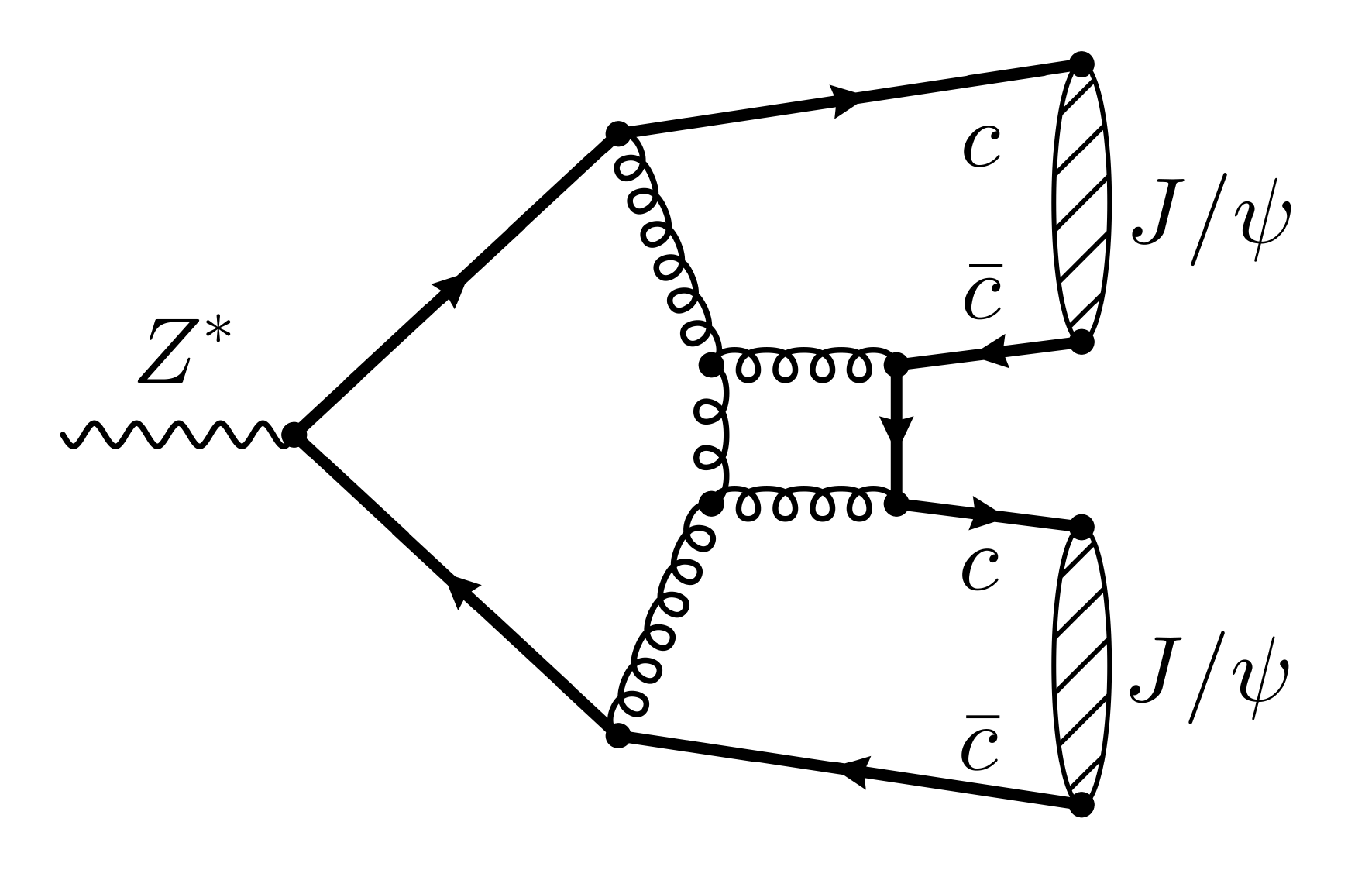}
    \subcaption{}
    \end{minipage}
    \begin{minipage}[b]{.3\linewidth}
        \centering
        \includegraphics[width=1.0\linewidth]{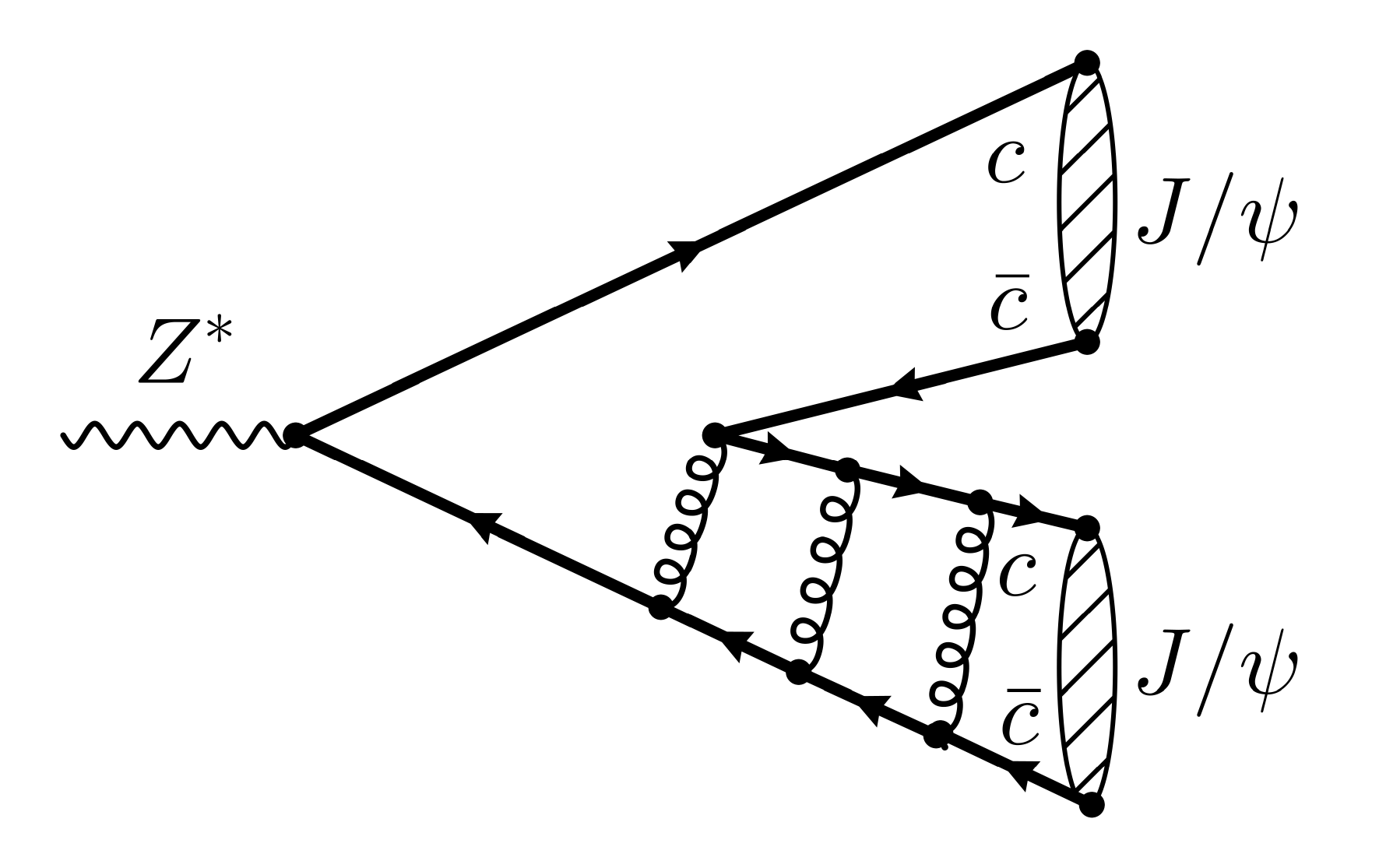}
    \subcaption{}
    \end{minipage}
    \begin{minipage}[b]{.3\linewidth}
        \centering
        \includegraphics[width=1.0\linewidth]{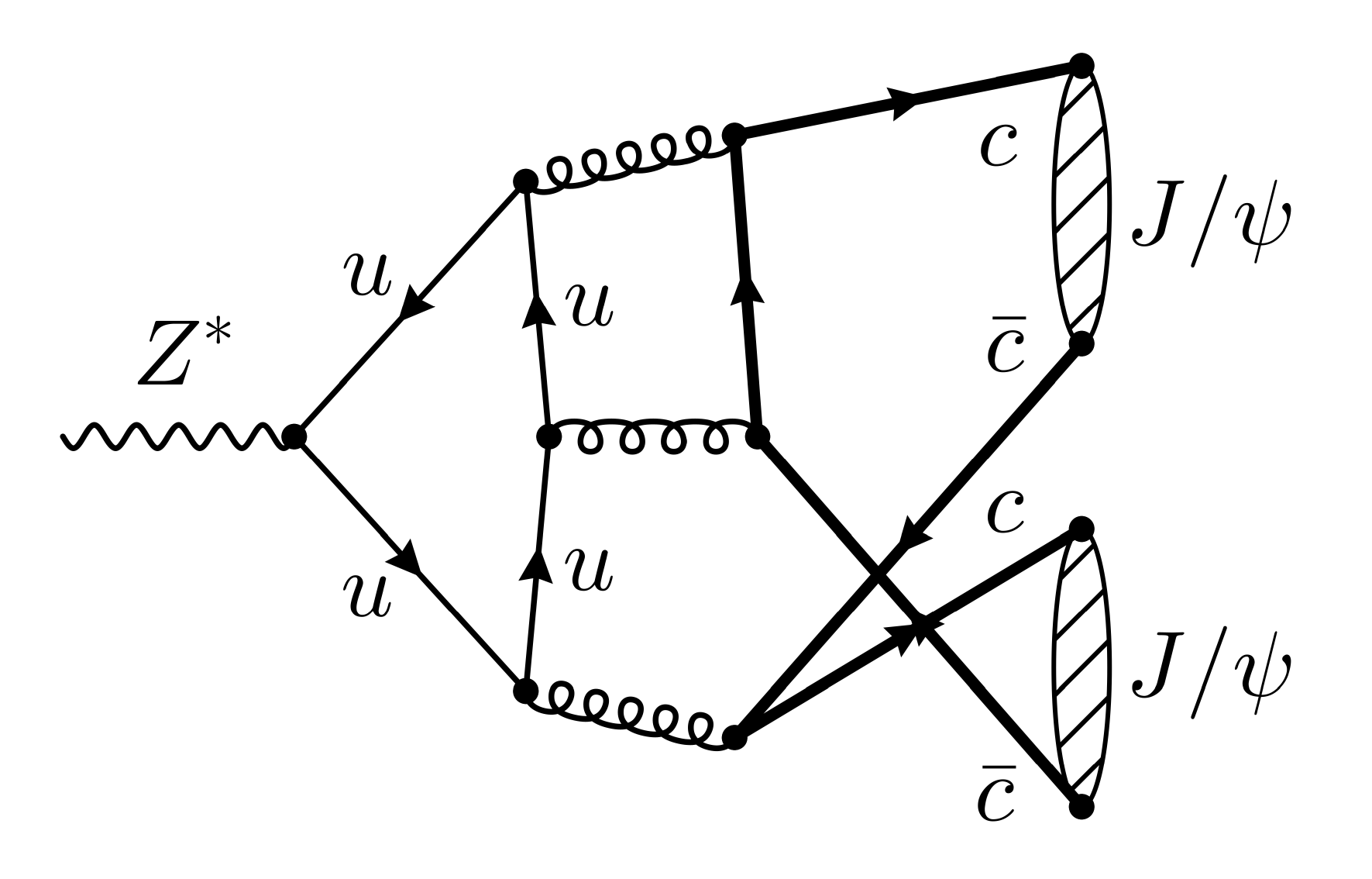}
    \subcaption{}
    \end{minipage}
    \begin{minipage}[b]{.3\linewidth}
        \centering
        \includegraphics[width=1.0\linewidth]{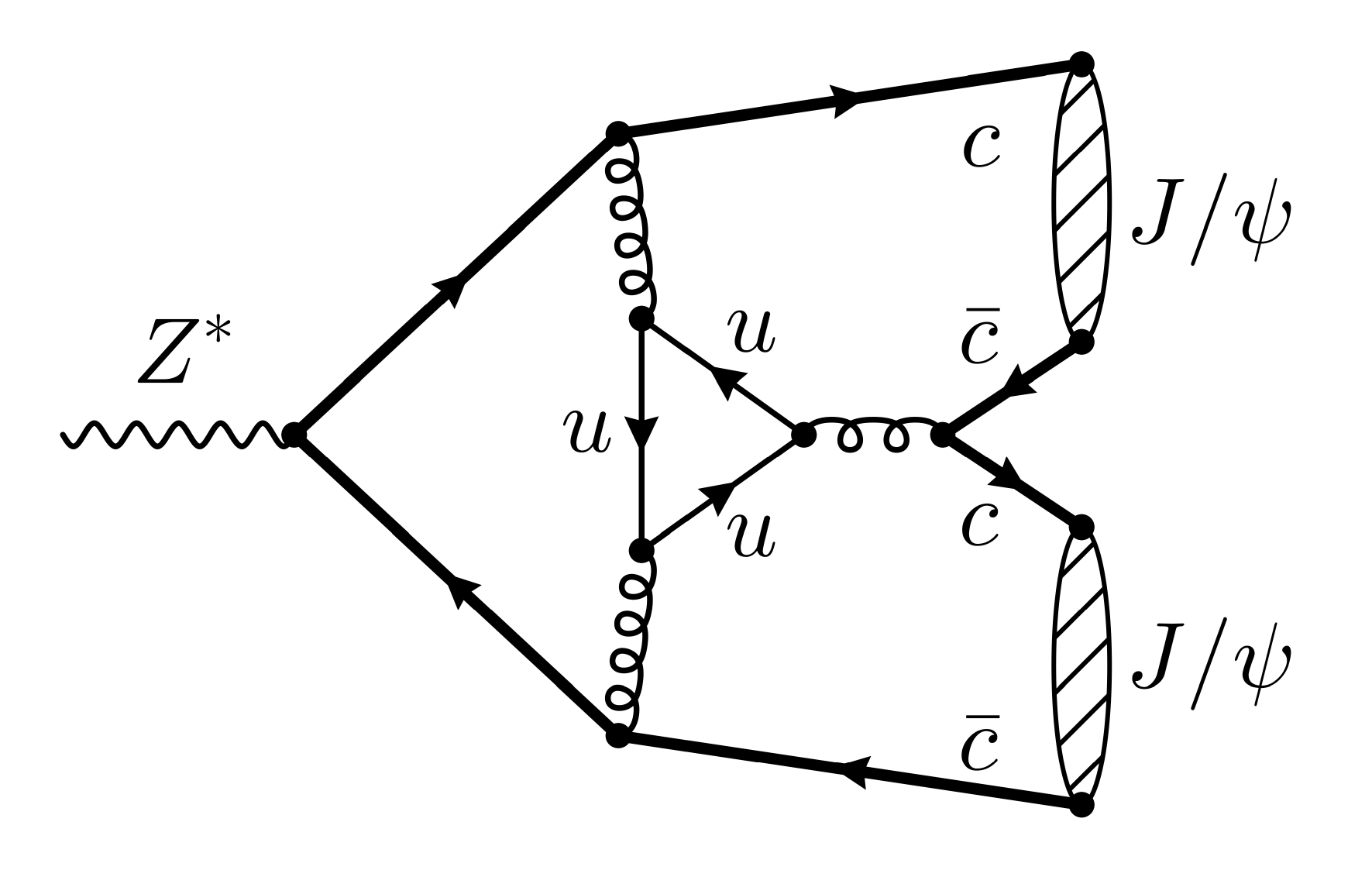}
    \subcaption{}
    \end{minipage}
        \begin{minipage}[b]{.3\linewidth}
        \centering
        \includegraphics[width=1.0\linewidth]{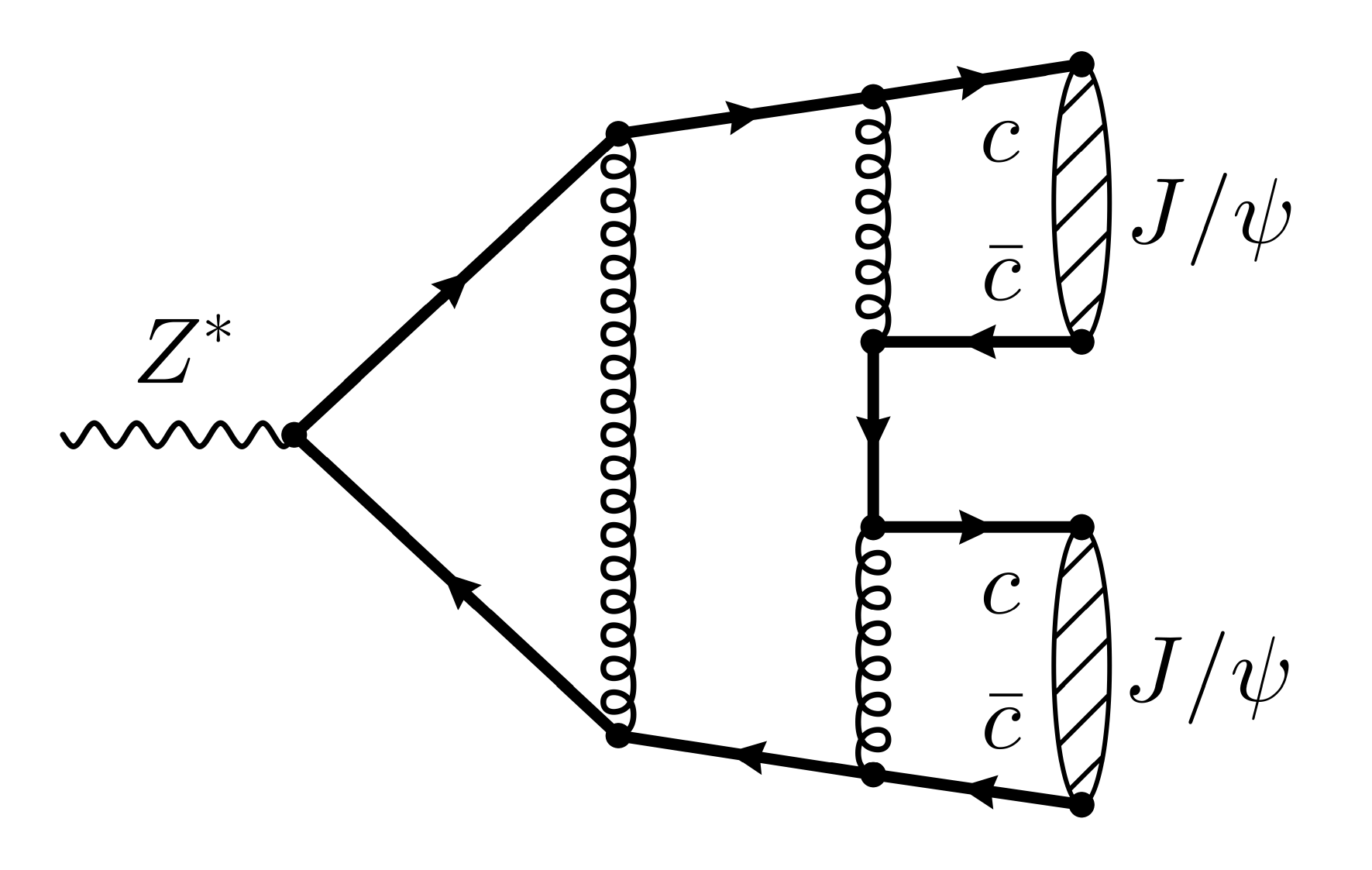}
    \subcaption{}
    \end{minipage}
    \begin{minipage}[b]{.3\linewidth}
        \centering
        \includegraphics[width=1.0\linewidth]{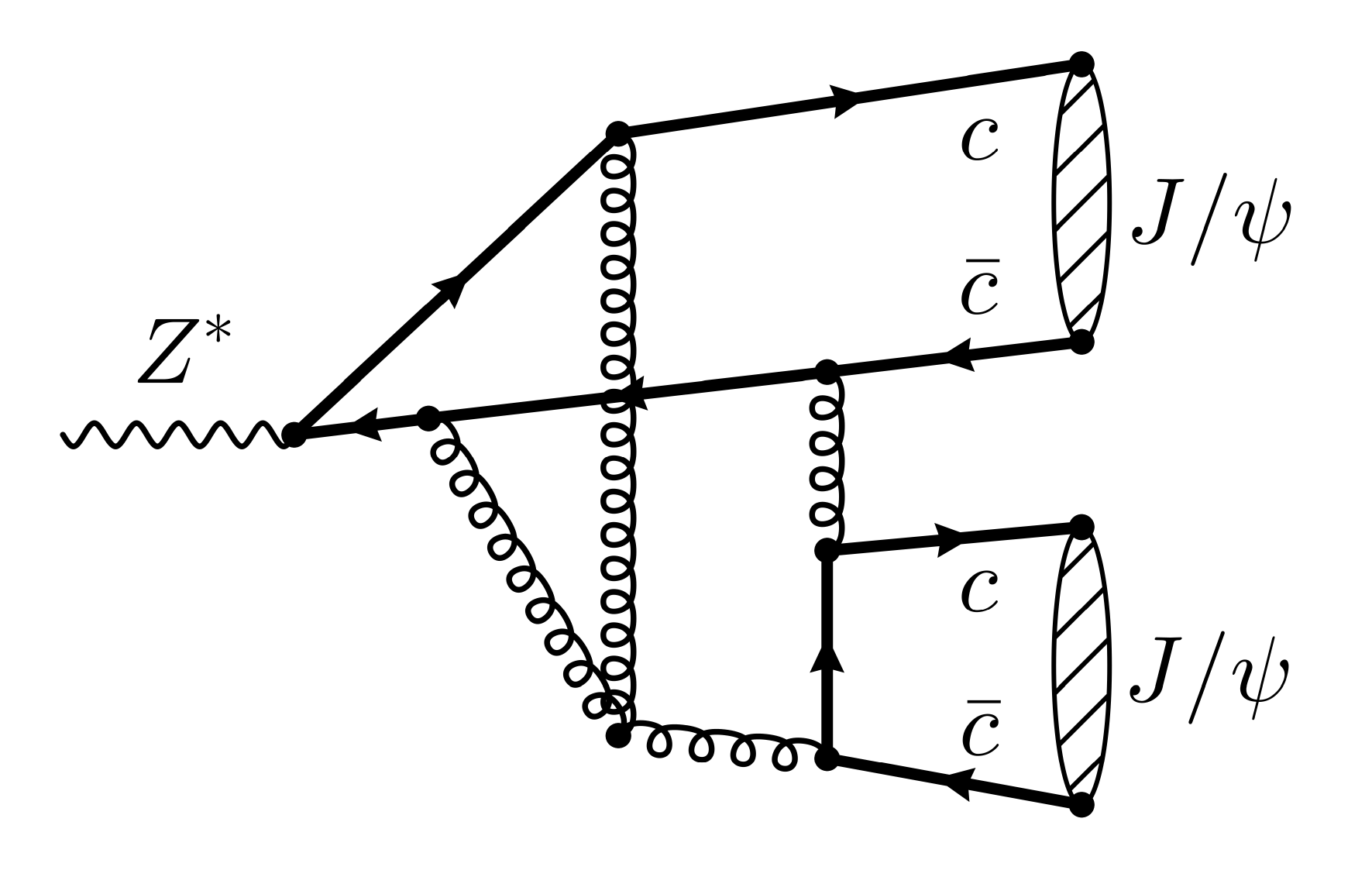}
    \subcaption{}
    \end{minipage}
    \caption{Some typical two-loop diagrams for  $Z \rightarrow J/\psi+J/\psi$.}\label{fig:ZJJloop2}
\end{figure*}

We generate the two-loop Feynman diagrams using {\tt FeynArts}~\cite{Hahn:2000} and {\tt QGRAF}~\cite{Nogueira:1991ex}, with sample diagrams shown in Fig.~\ref{fig:ZJJloop2}. Then we use the {\tt CalcLoop} package to deal with the Dirac and $SU(N_{c})$ color algebra ~\cite{www:CL}. The Feynman integrals are in general the Lorentz-tensor integrals, which are reduced to scalar integrals for subsequent processing. More than $10^4$ integrals are categorized into 153 families according to their propagators. 
The coefficients of these integrals are rational functions of the center-of-mass energy $s$, the charm quark mass $m_{c}$ as well as the dimensional regulator $\epsilon=(4-D)/2$, multiplied by the Lorentz tensors. Next, we utilize the package {\tt Blade}\cite{Guan:2024byi}, which is based on the method of block-triangular form\cite{Liu:2018dmc,Guan:2019bcx}, to perform integration-by-part (IBP) reduction\cite{Chetyrkin:1981qh}. This allows us to express integrals in each family as linear combinations of so-called master integrals(MIs).  The finite field method~\cite{vonManteuffel:2014ixa,Peraro:2016wsq, Peraro:2019svx, Klappert:2019emp, Klappert:2020aqs, Belitsky:2023qho} are employed in the reduction step.

Further, the master integrals are computed using the method of differential equations~\cite{Kotikov:1990kg} based on power series expansion~\cite{Caffo:2008aw, Czakon:2008zk,Lee:2014ioa,Moriello:2019yhu, Hidding:2020ytt, Armadillo:2022ugh}. The differential equations of master integrals with respect to $r=m_c^2/s$ are constructed using aforementioned IBP reduction. The boundary conditions at $r=1/100$ are obtained using the auxiliary mass flow method~\cite{Liu:2017jxz,Liu:2020kpc, Liu:2021wks, Liu:2022mfb} implemented in {\tt AMFlow} package~\cite{Liu:2022chg}.

The {\tt DEsolver} module in {\tt AMFlow}  solves the differential equations for each family, expressing all MIs as aymptotic expansions at $r=0$ (equivalently $s=\infty$). By combining the solutions of master integrals and the IBP reduction results, the amplitude can be expressed as an aymptotic expansion at $r=0$,
\begin{equation}\label{eq:asy0}
\mathcal{A}(\epsilon, r)=\sum_{\lambda, k, n} c_{\lambda, k, n}(\epsilon) r^{\lambda(\epsilon)} \ln ^k(r) r^n .
\end{equation}
Hereafter, the Lorentz index $\mu$ on the amplitude has been omitted for brevity.
Finally, the cross section or decay width is obtained by combining the two-loop, one-loop and Born amplitudes.

\section{Renormalization and IR Subtraction}
The bare amplitudes are renormalized according to
\begin{equation}
 \mathcal{A}(\alpha_s, m_Q)= Z_{2, Q}^{-2}Z_{A}Z_{5}\mathcal{A}_{0}(\alpha_{s,0}, m_{Q, 0}),
\end{equation}
where $Z_{2,Q}$ is the quark field renormalization constant and $Z_{A}$ and $Z_{5}$ will be explained later. The bare mass and the strong coupling constants are renormalized as follows:
\begin{equation}
\begin{aligned}
    &m_{Q, 0}=Z_{m, Q} m_Q, \\
    &\alpha_{s, 0}=(\frac{e^{\gamma_E}}{4 \pi})^\epsilon \mu_R^{2 \epsilon} Z_{\alpha_s}^{\overline{\mathrm{MS}}} \alpha_s(\mu_R).
\end{aligned}
\end{equation}

The field strength renormalization constant for gluon or light quark show up at intermediate stage in the two-loop calculation, but they completely cancel out in the results. 
The renormalization constants to $\mathcal{O}(\alpha_s^2)$ are taken from the Ref.~\cite{Barnreuther:2013qvf}.

In addition to the usual field, mass, and coupling constant renormalization, the axial current operator must be renormalized multiplicatively by a factor $Z_A$, and the correct Ward identities must be restored through a finite renormalization factor $Z_5$. Moreover, both $Z_A$ and $Z_5$ have distinct non-singlet and singlet versions. Here, ``non-singlet'' refers to the case where the Z boson couples to an external fermion line. In our calculation, the tree-level amplitudes involve only non-singlet contributions, whereas the one-loop amplitudes contain both non-singlet and singlet components. Consequently, we require $Z_{A,NS}$ and $Z_{5,NS}$ up to two loops, and $Z_{A,S}$ and $Z_{5,S}$  up to one loop. The explicit results are~\cite{Fael:2023zqr}:
\begin{equation}
\begin{aligned}
&Z_{A, NS}= 1+\left( \frac{\alpha_s}{\pi} \right)^2 \frac{1}{\epsilon} \left(\frac{11}{24} C_A C_F-\frac{1}{6} C_F T_F n_f \right),\\
& Z_{5, NS}=  1-\frac{\alpha_s}{\pi} C_F \\
&+\left(\frac{\alpha_s}{\pi} \right)^2\left(-\frac{107}{144} C_A C_F+\frac{11}{8} C_F^2+\frac{1}{36} C_F T_F n_f\right) \\
&Z_{A, S} = 1, \\
&Z_{5, S} = 1-\frac{\alpha_s}{\pi} C_F.
\end{aligned}
\end{equation}

After ultraviolet renormalization, a residual infrared pole remains. The infrared divergence can be factorized into the NRQCD matrix element $\left\langle\mathcal{O}^{J / \psi}\right\rangle$ under the $\overline{\mathrm{MS}}$ prescription. As a result, the two-loop amplitudes acquire a dependence on the NRQCD factorization scale $\mu_{\Lambda}$ :
\begin{equation}\label{eq:infrared}
\tilde{\mathcal{A}}^{(2l)}=\mathcal{A}^{(2l)}+\mathcal{A}^{(0l)} \alpha_s^2(\mu_R) \frac{\gamma_{J / \psi}}{\epsilon} \left(\frac{\mu_{\Lambda}^2 }{\mu_R^2}\right)^{-2 \epsilon},
\end{equation}
where the anomalous dimension for $J / \psi$ NRQCD current is given by~\cite{Beneke:1997jm,Czarnecki:2001zc}:
\begin{equation}
\gamma_{J / \psi}=-\frac{\pi^2}{12} C_F(2 C_F+3 C_A).
\end{equation}
After performing all the subtractions described above, we find that the two-loop amplitudes $\tilde{\mathcal{A}}^{(2l)}_{j}$ ($j=1,...,6$ as in Tab.~\ref{tab:ZJJbAmp}) for different Lorentz structures are individually free of divergence. The fact that each structure of finite provides a strong consistency check of our calculation. For brevity, we will omit the tilde symbol in what follows.

\section{Analytical Results}

For subsequent discussions, it is more convenient to work directly with the dimensionless functions $g_j(r)$, defined in Eq.~\eqref{eq:ME}. The leading-order results for $g_j~(j=1,...6)$ are summarized in Tab.~\ref{tab:ZJJbAmp}, where we have explicitly verified Eq.~\eqref{eq:GI} at this order. Although $g_4$ and $g_5$ vanish at this order, they become nonzero starting from next-to-leading order (NLO), with their leading contributions scaling as $\mathcal{O}(r^3)$. Owing to Eq.~\eqref{eq:GI}, only two form factors, $g_1$ and $g_3$, are indepenedent. Therefore, in the following, we present analytical results only for these two.

\begin{table*}[t]
\renewcommand\arraystretch{2}
{\begin{tabular}{c|c|c|c|c|c|c}
\hline Lorentz structures $b_j^{\alpha_1 \alpha_2 \mu}$         &$\epsilon^{\alpha_1 \alpha_2 \mu P_1}$              &$\epsilon^{\alpha_1 \alpha_2 \mu P_2}$             
    &$\epsilon^{\alpha_2 \mu P_1 P_2}  P_{1}^{\alpha_1}/m_c^2$         &$\epsilon^{\alpha_2 \mu P_1 P_2}  P_{2}^{\alpha_1}/m_c^2$         
    &$\epsilon^{\alpha_1 \mu P_1 P_2}  P_{1}^{\alpha_2}/m_c^2$         &$\epsilon^{\alpha_1 \mu P_1 P_2}  P_{2}^{\alpha_2}/m_c^2$ \\
\hline Form Factors $g^{(0l)}_{j}(r)$       &$512 i r^2$     &$-512 i r^2$            &$128 i r^2$    &0    &0     &$-128 i r^2$  \\
\hline
\end{tabular}
}
\caption{Reduced Born amplitudes.}\label{tab:ZJJbAmp}
\end{table*}

The functions $g_j(r)$ admit the following asymptotic expansion in the variable $r$:
\begin{equation}\label{eq:JJExpand}
\begin{aligned}
g_{j}^{(0l)}(r)=& r^2 G_{j,(2,0)}^{(0l)}  \\
g_{j}^{(1l)}(r)=& r^2 \left[G_{j,(2,2)}^{(1l)} \;  \ln^2(r) +G_{j,(2,1)}^{(1l)} \;  \ln(r)  +G_{j,(2,0)}^{(1l)}\right] \\
&+\mathcal{O}(r^3) \\
g_{j}^{(2l)}(r)=&r^2 \left[G_{j,(2,4)}^{(2l)}  \;  \ln^4(r) +G_{j,(2,3)}^{(2l)} \; \ln^3(r) \right. \\
&\left. + G_{j,(2,2)}^{(2l)} \; \ln^2(r) +G_{j,(2,1)}^{(2l)} \; \ln(r) +G_{j,(2,0)}^{(2l)} \right] \\
&+\mathcal{O}(r^3)
\end{aligned}
\end{equation}

The analytical expressions for  $G_{j,(2,k)}^{(il)}$ constitute the principal findings of this work, which are reconstructed from high-precision numerical coefficients using the PSLQ algorithm, with the basis of irrational numbers detailed in Ref.~\cite{Chen:2025qgy}.
The NLO results are given by
\begin{equation}\label{eq:NLO1}
\begin{aligned}
g_{1}^{(1l)}(r)/r^2=&\frac{32 i \ln ^2(r)}{\pi^2}-\frac{32 i(-33-18 i \pi-120 \ln (2)) \ln (r)}{9 \pi^2} \\
&-\frac{32 i}{9 \pi^2}\left(-75 \ln (\mu^2)-120 \ln ^2(2) -120 i \pi \ln (2) \right.\\
&\left.-306 \ln (2)+11 \pi^2-33 i \pi+19\right),
\end{aligned}
\end{equation}
\begin{equation}\label{eq:NLO3}
\begin{aligned}
g_{3}^{(1l)}(r)/r^2=&\frac{24 i \ln ^2(r)}{\pi^2}+\frac{8(-63 i-54 \pi+330 i \ln (2)) \ln (r)}{9 \pi^2} \\
&+\frac{8}{9 \pi^2}\left(75 i \ln (\mu^2)+495 i \ln ^2(2)-330 \pi \ln (2) \right.\\
&\left.+602 i \ln (2)-28 i \pi^2+63 \pi-69 i\right).
\end{aligned}
\end{equation}

The complete expressions for two-loop amplitudes are rather lengthy so that we present only the leading and next-to-leading logarithmic terms in Eq.(\ref{eq:JJExpand}). For $j=1$,
\begin{equation}
\begin{aligned}
&G_{1,(2,4)}^{(2l)}=\frac{4 i}{9 \pi^4}, \\
&G_{1,(2,3)}^{(2l)}=-\frac{4(-69 i+12 \pi-134 i \ln (2))}{27 \pi^4}.
\end{aligned}
\end{equation}
For $j=3$,
\begin{equation}
\begin{aligned}
&G_{3,(2,4)}^{(2l)}=\frac{7 i}{27 \pi^4}, \\
&G_{3,(2,3)}^{(2l)}=-\frac{-90 i+56 \pi-991 i \ln (2)}{54 \pi^4}.
\end{aligned}
\end{equation}
The full analytical results at $\mathcal{O}(r^2)$ including all logarithmic terms are provided in the supplementary file.

\section{Numerical Results}
In this section, we will discuss the phenomenological prediction. The parameters we used in the computation are outlined in Tab.~\ref{tab:numeric}. The fine
structure constant is fixed in the Thomson limit $\alpha=1/137$. The strong coupling constant is evaluated to two loops accuracy with our in-house program, wherein $n_f=5$
for $\mu_{R} \geq 4.8 \text{GeV}$ while $n_f= 4$ for $\mu_{R} < 4.8 \text{GeV}$. 
\begin{table}[htb]
\renewcommand\arraystretch{2}
\centering{
\begin{tabular}{cc}
\hline              
$m_c=1.5 \text{GeV}\qquad $          & $R_{J/\psi}^2=1.1 \text{GeV}^3$ \\
\hline
$m_b=4.8 \text{GeV}\qquad $          & $R_{\Upsilon}^2=6.477 \text{GeV}^3$ \\
\hline
$m_Z=91.2 \text{GeV}\qquad $         & $\Gamma_Z=2.5 \text{GeV}$ \\
\hline
$\sin\theta_w=38/79\qquad $          & $\cos\theta_w=64/73$ \\
\hline
\end{tabular}
}
\caption{Numerical values for parameters.}\label{tab:numeric}
\end{table}

\subsection{Cross section of $J /\psi + J/ \psi$ production}
\begin{figure*}[htb]
	\centering
\begin{minipage}[b]{.4\linewidth}
\centering{
  \includegraphics[width=1.0\linewidth]{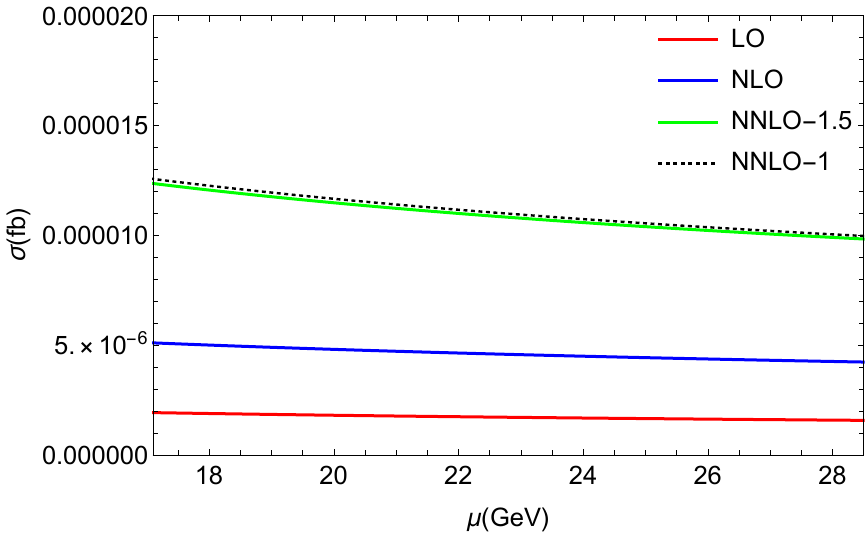}
  \subcaption{ $\sqrt{s}=m_Z/4=22.8\text{GeV}$.}\label{fig:csJJp25}
  }
\end{minipage}
\begin{minipage}[b]{.4\linewidth}
\centering{
  \includegraphics[width=1.0\linewidth]{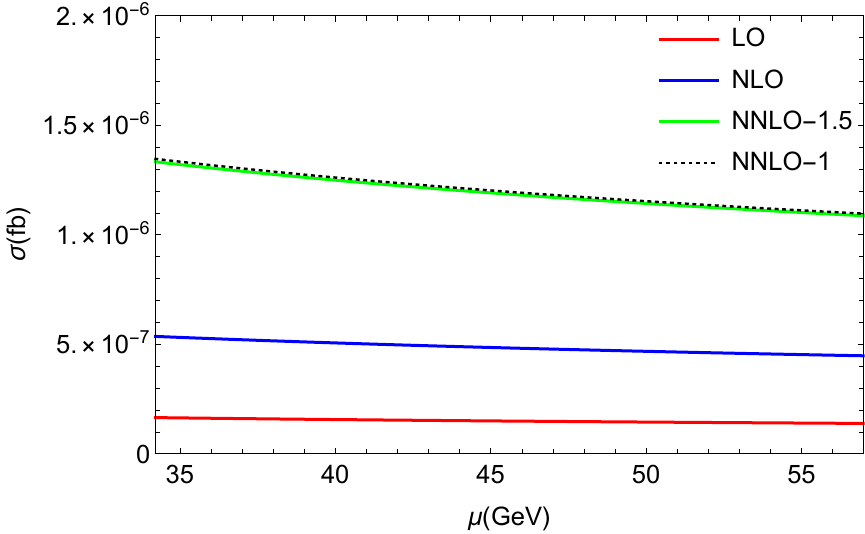}
  \subcaption{$\sqrt{s}=m_Z/2=45.6\text{GeV}$. }\label{fig:csJJp5}
  }
\end{minipage}
\begin{minipage}[b]{.4\linewidth}
\centering{
  \includegraphics[width=1.0\linewidth]{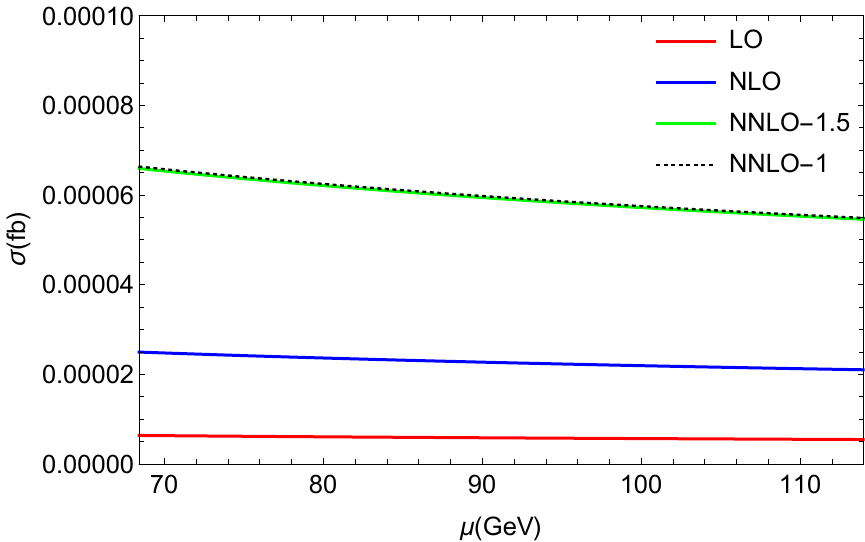}
  \subcaption{ $\sqrt{s}=m_Z=91.2\text{GeV}$. }\label{fig:csJJ1}
  }
\end{minipage}
\begin{minipage}[b]{.4\linewidth}
\centering{
  \includegraphics[width=1.0\linewidth]{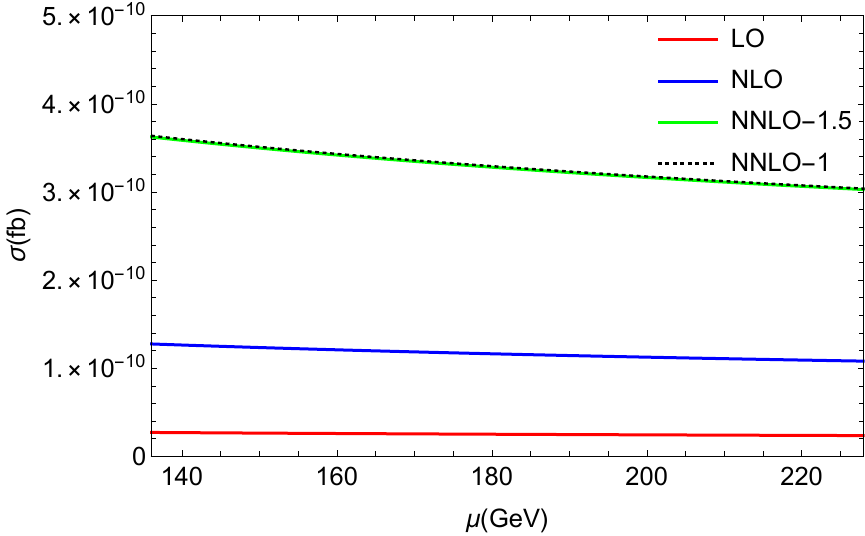}
  \subcaption{ $\sqrt{s}=2m_Z=182.4\text{GeV}$. }\label{fig:csJJ2}
  }
\end{minipage}
\caption{The cross section with respect to renormalization scale by varying $\mu_R$ for $\pm \sqrt{s}/4$.}\label{fig:csJJ}
\end{figure*}

In Tab.~\ref{tab:JJ}, we enumerate our predictions at various perturbative accuracies for several colliding energy of interest, including $\sqrt{s}=m_Z/4,\: m_Z/2,\:m_Z,\: 2\,m_Z$ as well as $\sqrt{s}=10.52\text{GeV},\: 10.58\text{GeV},\: 250\text{GeV}$. The center value correspond to $\mu_{R}=\sqrt{s}$, with the first uncertainty estimated by varying  $\mu_R$ between $\sqrt{s}/2$ and $2\sqrt{s}$, and the second uncertainty obtained by varying $m_c$ within $1.5 \pm 0.2$~GeV.  An overall comparison indicates that the $e^{+} e^{-} \rightarrow Z^{\ast} \rightarrow J /\psi + J/ \psi$ production cross section is relatively small compared to those of other channels, such as $\sigma(e^{+} e^{-} \rightarrow \gamma\gamma \rightarrow J /\psi+J/ \psi)$ or $\sigma(e^{+} e^{-} \rightarrow \gamma \rightarrow J /\psi+\eta_c)$, etc~\cite{Chen:2025qgy,Sang:2023liy}. The uncertainty from $\mu_R$ dependence can be estimated. Taking $\sqrt{s}=45.6\text{GeV}$ as an example, the differences of finite cross section at $\mu_{R}=(\sqrt{s}/2,\;2\sqrt{s})$ from $\mu_{R}=\sqrt{s}$ are:
\begin{center}
\begin{math}
\begin{aligned}
    &(+28.9\%,\;-20.0\%) \quad\text{for} \quad\sigma_{LO}, \\
    &(+29.9\%,\;-20.5\%) \quad\text{for} \quad\sigma_{NLO} ,\\
    &(+34.8\%,\;-22.6\%) \quad\text{for} \quad\sigma_{NNLO}. 
\end{aligned}
\end{math}
\end{center}
That is to say, the $\mu_R$ dependence become slightly heavier in higher $\alpha_s$ order, as illustrated soon in Fig.\ref{fig:csJJ}.

Fig.~\ref{fig:csJJp25}–\ref{fig:csJJ2} display the cross section as a function of $\mu_R$ for $\sqrt{s}=m_Z/4, \: m_Z/2, m_Z, \: 2\,m_Z$ respectively.
The absolute magnitudes vary significantly among these energy points, up to several orders. Nevertheless, these figures show similar patterns: The NLO cross section is 3 times larger than LO, and the NNLO cross section is 2 times larger than NLO.  In general, the relative size can be summarized as following:
\begin{equation}\label{eq:K2factor}
\frac{\sigma_{N N L O}}{ \sigma_{N L O}} \approx 2 \sim 3.
\end{equation}
We can also estimate the K factor, which is consistent to the Ref.~\cite{Berezhnoy:2021tqb}:
\begin{equation}
\frac{\sigma_{N L O} }{ \sigma_{L O}} \approx 3 \sim 5.
\end{equation}

Finally, we check the dependence on the factorization scale $\mu_{\Lambda}$.
For example, at the colliding energy $\sqrt{s}=45.6\text{GeV}$, taking the renomalization scale $\mu_{R}=45.6\text{GeV}$ and the factorization scale, as in Eq.(\ref{eq:infrared}), $\mu_{\Lambda}=1.5\text{GeV}$, we have $\sigma_{NNLO}=1.184\times 10^{-6} $fb.
If we change $\mu_{\Lambda}$ to $1.0\text{GeV}$, we have $\sigma_{NNLO}=1.195\times 10^{-6} $fb.
In Fig.~\ref{fig:csJJ}, the solid line ``NNLO-1.5" and the dotted line ``NNLO-1"  correspond to $\mu_{\Lambda}=1.5\text{GeV}$ and $\mu_{\Lambda}=1\text{GeV}$ respectively.
It is evident that the $\mu_{\Lambda}$ dependence is very modest compared to the $\mu_{R}$ dependence. For this reason, we will fix $\mu_{\Lambda}=1.5\text{GeV}$ in the following discussions. 

\begin{table*}[htb]
\renewcommand\arraystretch{2}
\setlength{\belowcaptionskip}{0.2cm}
\centering{
\begin{tabular}{c|c|c|c}
\hline
$\sqrt{s}$/ GeV        & LO/ fb     & NLO/ fb   & NNLO/ fb  \\
\hline
10.52&     $2.128_{-0.919-0.697}^{+0.551+0.656}\times 10^{-5}$ & $4.620_{-1.51-1.86}^{+1.01+1.57}\times 10^{-5}$ & $11.50_{-6.45-4.72}^{+3.28+3.91}\times 10^{-5}$ \\
\hline
10.58&     $2.103_{-0.906-0.676}^{+0.544+0.639}\times 10^{-5}$ & $4.574_{-1.49-1.82}^{+0.998+1.54}\times 10^{-5}$ & $11.37_{-6.36-4.60}^{+3.24+3.82}\times 10^{-5}$ \\
\hline
22.8&      $1.714_{-0.586-0.0805}^{+0.384+0.0893}\times 10^{-6}$ & $4.578_{-1.43-0.500}^{+0.965+0.449}\times 10^{-6}$ & $10.82_{-4.40-1.50}^{+2.65+1.26}\times 10^{-6}$ \\
\hline
45.6&      $1.487_{-0.429-0.0164}^{+0.298+0.0186}\times 10^{-7}$ & $4.824_{-1.44-0.338}^{+0.988+0.291}\times 10^{-7}$ & $11.84_{-0.12-1.23}^{+2.68+1.01}\times 10^{-7}$ \\
\hline
91.2&      $0.5759_{-0.144-0.00156}^{+0.104+0.00178}\times 10^{-5}$ & $2.256_{-0.637-0.131}^{+0.446+0.111}\times 10^{-5}$ & $5.911_{-1.86-0.543}^{+1.26+0.443}\times 10^{-5}$ \\
\hline
182.4&     $0.2481_{-0.0548-0.00017}^{+0.0410+0.00019}\times 10^{-10}$ & $1.157_{-0.306-0.0594}^{+0.220+0.0504}\times 10^{-10}$ & $3.265_{-0.947-0.275}^{+0.665+0.226}\times 10^{-10}$ \\
\hline
250&     $1.372_{-0.288-0.000492}^{+0.218+0.000562}\times 10^{-11}$ & $6.892_{-1.77-0.338}^{+1.28+0.287}\times 10^{-11}$ & $20.12_{-5.65-1.64}^{+4.01+1.34}\times 10^{-11}$ \\
\hline
\end{tabular}
}
\caption{The cross sections for $e^+e^- \to Z^{\ast} \to J/\psi + J/\psi$ at different colliding energies. The central values are evaluated at $\mu_R = \sqrt{s}$ and $m_c = 1.5$~GeV. The first uncertainty is estimated by varying $\mu_R$ from $\sqrt{s}/2$ to $2\sqrt{s}$, and the second estimated by varying $m_c$ from $1.3$~GeV to $1.7$~GeV.The factorization scale $\mu_\Lambda=m_c=1.5~\text{GeV}$.}\label{tab:JJ}
\end{table*}

\begin{figure}[htb]
\centering{
  \includegraphics[width=0.96\linewidth]{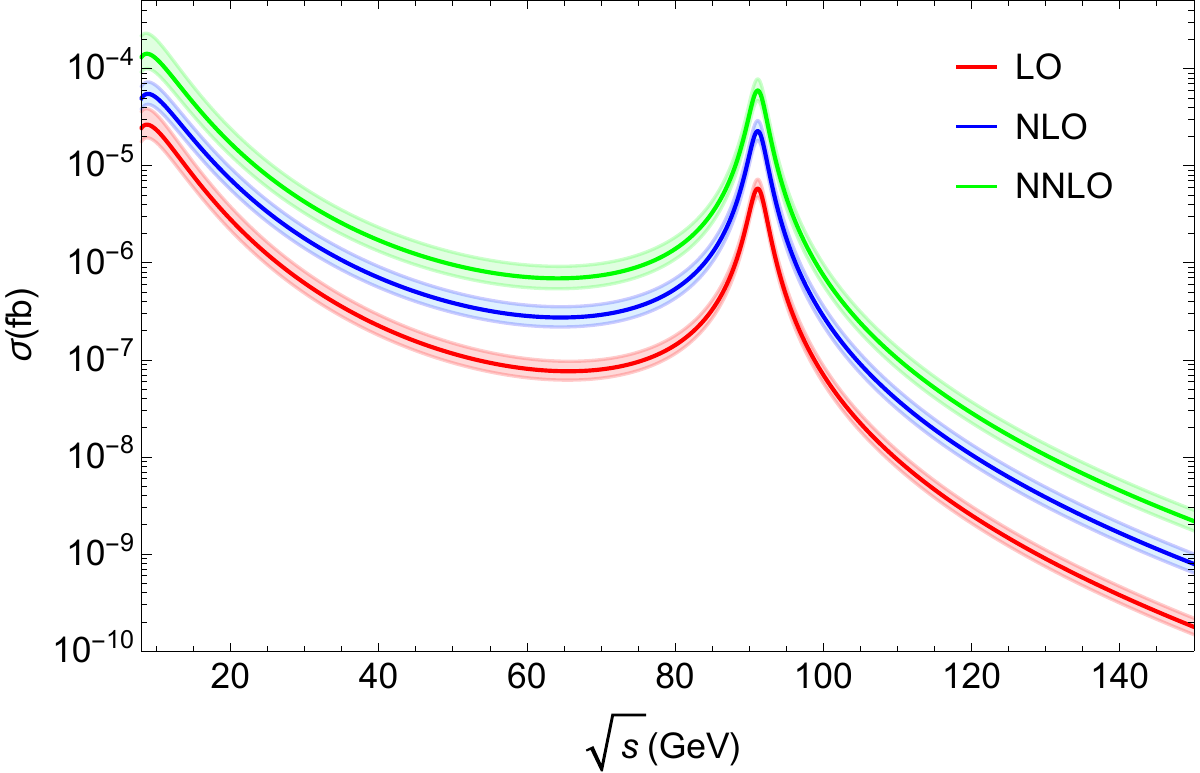}
  \caption{Cross section of $J /\psi+ J/ \psi$ production as a function of center-of-mass energy. The green,blue,red bands correspond to the NNLO, NLO and LO prediction respectively. Width of the bands represent theoretical uncertainties by varying $\mu_R$ from $\sqrt{s}/2$ to $2\sqrt{s}$. The factorization scale is fixed at $\mu_{\Lambda} = 1.5~\text{GeV}$. The vertical axis uses a logarithmic scale. } \label{fig:JJ}
  }
\end{figure}

In Fig.~\ref{fig:JJ}, the cross sections  for $J/\psi + J/\psi$ production are depicted at the range of $\sqrt{s}=[8,150]$~GeV, with the red, blue, and green bands stand for the LO, NLO, and NNLO cross sections, respectively.  Each line are three fold and the band width represent the theoretical uncertainty from the renormalization scale variation: the upper one stands for $\mu_{R}=\sqrt{s}/2$, the centered one for $\mu_{R}=\sqrt{s}$, and the lower one for $\mu_{R}=2\sqrt{s}$.
The cross section displays a prominent peak at the $Z$-boson pole ($\sqrt{s} = 91.2$GeV), which is of particular interest for future lepton colliders. Away from the resonance, the cross section decreases rapidly.

At the collision energy $\sqrt{s}=91.2\text{GeV}$, there is a significant enhancement in the cross section. Alternatively, one may interpret this as $\sigma(\sqrt{s})$ being suppressed by the large mass of the Z boson when $\sqrt{s}<m_Z$, while the cross section at $\sqrt{s}=91.2\text{GeV}$ is not subject to such suppression. Specifically, we find that

\begin{equation}
\sigma(91.2~\text{GeV})=5.911_{-1.86-0.543}^{+1.26+0.443}\times 10^{-5} ~\text{fb},
\end{equation}
which is sufficiently large to be observed in forthcoming Z factories \cite{CEPCStudyGroup:2018ghi,ILC:2013jhg}. 

In summary, our findings are crucial for elucidating the complexities associated with the perturbative convergence in QCD and its effective theory~\cite{Chen:2025qgy}.

\subsection{Cross section of $\Upsilon + \Upsilon$ production}
\begin{figure}[htb]
\centering{
  \includegraphics[width=0.96\linewidth]{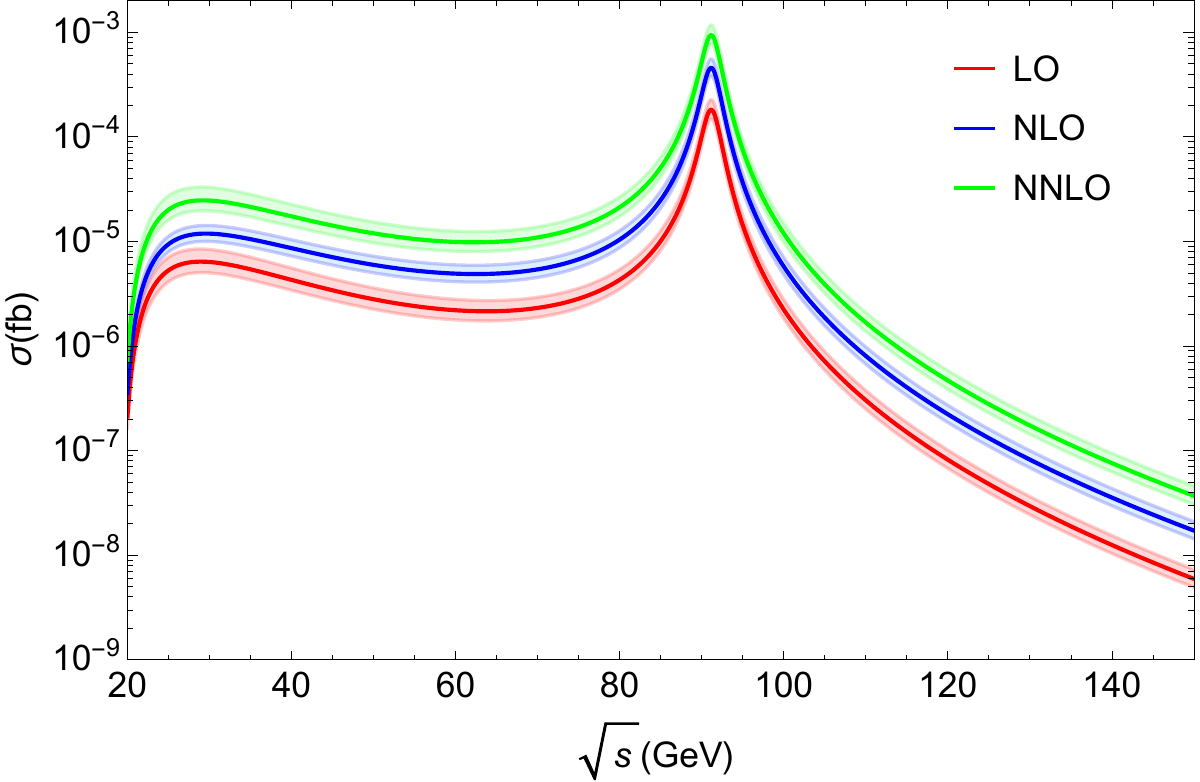}
  \caption{Cross section of $\Upsilon + \Upsilon$ production as a function of center-of-mass energy. The green,blue,red bands correspond to the NNLO, NLO and LO prediction respectively. Width of the bands represent theoretical uncertainties by varying $\mu_R$ from $\sqrt{s}/2$ to $2\sqrt{s}$. The factorization scale is fixed at $\mu_{\Lambda} = 4.8~\text{GeV}$. The vertical axis uses a logarithmic scale.} \label{fig:UU}
  }
\end{figure}

As a nontrivial byproduct, our result can be easily utilized for the study of double bottomonium production. Here we simply provide our numerical prediction in Tab.\ref{tab:UU} and Fig.\ref{fig:UU}. 
In Fig.\ref{fig:UU}, the cross sections  for $\Upsilon + \Upsilon$ production are depicted at the range of $\sqrt{s}=[20,150]$~GeV. The plot style is similar to Fig.\ref{fig:JJ}.
From Tab.\ref{tab:UU}, we can read the peaking value at 91.2GeV is 
\begin{equation}
\sigma(91.2~\text{GeV})=2.049_{-0.643-0.188}^{+0.438+0.154}\times 10^{-3} ~\text{fb},
\end{equation}
which is one order larger than its charmonium counterpart. 

\begin{table*}[htb]
\renewcommand\arraystretch{2}
\setlength{\belowcaptionskip}{0.2cm}
\centering{
\begin{tabular}{c|c|c|c}
\hline$\sqrt{s}$/ GeV        & LO/ fb     & NLO/ fb   & NNLO/ fb  \\
\hline22.8& $0.594_{-0.203-0.0279}^{+0.133+0.0309} \times 10^{-4}$ & $1.587_{-0.497-0.173}^{+0.335+0.156}\times 10^{-4}$ & $3.751_{-1.53-0.518}^{+0.92+0.438}\times 10^{-4}$ \\
\hline45.6&  $0.516_{-0.149-0.00567}^{+0.103+0.00644}\times 10^{-5}$ & $1.673_{-0.500-0.117}^{+0.343+0.101\times 10^{-5}}$ & $4.107_{-1.43-0.426}^{+0.928+0.349}\times 10^{-5}$ \\
\hline91.2& $1.997_{-0.500-0.00540}^{+0.361+0.00617}\times 10^{-4}$ & $7.821_{-2.21-0.453}^{+1.55+0.384}\times 10^{-4}$ & $20.49_{-6.43-1.88}^{+4.38+1.54}\times 10^{-4}$ \\
\hline182.4& $0.860_{-0.190-0.000580}^{+0.142+0.000662}\times 10^{-9}$ & $4.012_{-1.06-0.206}^{+0.761+0.175}\times 10^{-9}$ & $11.32_{-3.28-0.955}^{+2.30+0.782}\times 10^{-9}$ \\
\hline250&  $0.476_{-0.0999-0.000171}^{+0.0757+0.000195}\times 10^{-10}$ & $2.389_{-0.613-0.117}^{+0.444+0.095}\times 10^{-10}$ & $6.977_{-1.96-0.567}^{+1.39+0.466}\times 10^{-10}$ \\
\hline
\end{tabular}
}
\caption{The cross sections for $e^+e^- \to Z^{\ast} \to \Upsilon + \Upsilon$ at different colliding energies. The central values are evaluated at $\mu_R = \sqrt{s}$ and $m_b = 4.8$~GeV. The first uncertainty is estimated by varying $\mu_R$ from $\sqrt{s}/2$ to $2\sqrt{s}$, and the second estimated by varying $m_c$ from $4.6$~GeV to $5.0$~GeV.The factorization scale $\mu_\Lambda=m_b=4.8~\text{GeV}$.}\label{tab:UU}
\end{table*}

\subsection{Rare decay of Z boson}
In the upcoming high luminosity LHC and the future super Z factory, significant amount of Z bosons will be produced, providing a good chance to detect many rared decay channels~\cite{Wang:2023ssg}. This offer a good oppotunity to investigate the Z property. In this subsection, we utilize Eq.(\ref{eq:DW}) to compute the decay width. Our results are shown in Tab.\ref{tab:DW}. Again, we use the numerical value in Tab.\ref{tab:numeric},  select $\mu_R=m_Z,\mu_{\Lambda}=m_Q$, and vary $\mu_R$ from $m_Z/2$ to $2m_Z$; vary $m_Q$ by $\pm 0.2$GeV . In conclusion, we have 
\begin{equation}
\begin{aligned}
 \mathcal{B}_{\text {QCD}}^{NNLO}(J/\psi+J/\psi)&=(2.139_{-0.672-0.197}^{+0.457+0.160}) \times 10^{-12},\\
 \mathcal{B}_{\text {QCD}}^{NNLO}(\Upsilon+\Upsilon)&=(3.371_{-0.850-0.114}^{+0.600+0.108}) \times 10^{-11}.
\end{aligned}
\end{equation}

In Ref.\cite{Gao:2022mwa,Li:2023tzx}, the authors found those QED diagrams with a photon directly fragmenting to a $J/\psi$ contribute the most to the $Z \rightarrow J/\psi+J/\psi$.
Recently, a theoretical advancement has been made regarding the NLO-QCD correction for both QCD-tree QED-tree diagrams\cite{Li:2023tzx}. The branch ratio of the Z decay to double $J/\psi$ by their calculation is:
\begin{equation}
\mathcal{B}_{\text {QED,QCD}}^{NLO}=(1.110_{-0.241-0.001}^{+0.334+0.054}) \times 10^{-10}.
\end{equation}
Applying analogous reasoning as Eq.(\ref{eq:K2factor}), it seems to be a justifiable inference that the NNLO correction to both QED-tree and QCD-tree diagrams for $Z \rightarrow  J/\psi+J/\psi$ would suggest:
\begin{equation}
\mathcal{B}_{\text {QED,QCD}}^{NNLO}\sim 3 \times 10^{-10}.
\end{equation}
We propose, the full NNLO-QCD correction to photon-fragmenting QED tree diagrams would be important for future precision study of the $Z \rightarrow J/\psi+J/\psi$  channels.

\begin{table*}[htb]
\renewcommand\arraystretch{2}
\setlength{\belowcaptionskip}{0.2cm}
\centering{
\begin{tabular}{c|c|c|c}
\hline
~      & LO/ GeV     & NLO/ GeV   & NNLO/ GeV  \\
\hline
$\Gamma_{J/\psi + J/\psi}$&     $0.520_{-0.130-0.0014}^{+0.094+0.0016}\times 10^{-12}$&     $ 2.037_{-0.575-0.118}^{+0.403+0.100}\times 10^{-12}$&     $ 5.338_{-1.68-0.491}^{+1.14+0.400}\times 10^{-12}$ \\
\hline
$\Gamma_{\Upsilon + \Upsilon}$&     $1.628_{-0.407-0.0154}^{+0.295+0.0160}\times 10^{-11}$&     $ 4.094_{-0.892-0.105}^{+0.670+0.101}\times 10^{-11}$&     $ 8.413_{-2.12-0.285}^{+1.50+0.270}\times 10^{-11}$ \\
\hline
\end{tabular}
}
\caption{The partial decay width for Z boson. The central values are evaluated at $\mu_R = m_Z=91.2$~GeV and $m_c = 1.5$~GeV or $m_b = 4.8$~GeV. The first uncertainty is estimated by varying $\mu_R$ from $m_Z/2$ to $2m_Z$, and the second from varying $m_{c,b}$ for $\pm 0.2$~GeV. The factorization scale is set as $\mu_\Lambda=m_{c,b}$, respectively.}\label{tab:DW}
\end{table*}

\section{Summary}
We investigate the exclusive double charmonium production in electron-positron annihilation, focusing on the process $e^{+} e^{-} \rightarrow Z^{\ast} \rightarrow J / \psi+J / \psi$. Our study provides new insights into rare Z boson decays. We carried out extensive calculations at next-to-leading order (NLO) and next-to-next-to-leading order (NNLO), employing various methods to ensure the consistency and reliability of our results.

To validate the treatment of $\gamma^5$ in the calculation, we employed both the Larin scheme and the cyclic scheme, finding excellent agreement between the two. 
The Feynman integral reductions were cross-checked against an independent code that combines {\tt FiniteFlow}~\cite{Peraro:2019svx} and {\tt LiteRed}~\cite{Lee:2012cn}.
All master integrals, obtained by solving differential equations with respect to $r$ with boundary conditions at $r = 1/100$, were verified by AMFlow at another phase space point $r = 1.5^2/91.2^2$, achieving agreement up to 10 significant digits.
Furthermore, the ultraviolet and infrared divergences  were systematically subtracted, providing strong consistency checks.

We have obtained, for the first time, the NNLO QCD corrections to $e^{+} e^{-} \rightarrow Z^{\ast} \rightarrow J/ \psi + J/ \psi$ and found that the NNLO correction is 2-3 times larger than the NLO correction. This result poses significant challenges to the perturbative convergence of QCD and underscores the phenomenological relevance of higher-order effects. 
Furthermore, the analytical asymptotic expansions derived in this work have direct implications for studies of the 4-lepton decay channel of the Z boson conducted in LHC \cite{CMS:2019wch,CMS:2022fsq,LHCb:2020bwg}, once combined with the leptonic decays of the $J/\psi$. Importantly, our results are independent of the Z boson production mechanism, thereby offering broader applicability to various experimental settings.

\begin{acknowledgments}
The authors would like to thank X.Liu for useful discussion. C-Q.He wishes to thank the warm hospitality extended to him by QCD Theory Group in UCLA.
This work is supported in part by the National Natural Science Foundation of
China (Grants  No. 12325503, No. 12035007), the High-performance Computing Platform of Peking University. XG is supported by the United States Department of Energy, Contract DE-AC02-76SF00515. XC is supported by the Swiss National Science Foundation (SNF) under contract 200020\_219367 and the UZH Postdoc Grant, grant no. [FK-25-104].
The Feynman diagrams in this paper are drawn with the aid of \texttt{FeynGame}~\cite{Bundgen:2025utt}.
\end{acknowledgments}

\bibliography{refs}

\end{document}